\documentclass[prb,twocolumn,superscriptaddress,amsmath,amssymb,showpacs,floatfix,preprintnumbers]{revtex4-1}
\usepackage{amsmath}
\usepackage{amssymb}
\usepackage{graphicx}
\usepackage[export]{adjustbox}
\usepackage{epstopdf}
\usepackage{color}
\usepackage{bm}
\usepackage[english]{babel}
\usepackage{epsfig}
\usepackage{amsmath}
\usepackage[colorlinks=true,citecolor=blue,linkcolor=blue]{hyperref}
\usepackage{amssymb}

\DeclareMathAlphabet{\bi}{OML}{cmm}{b}{it}

\begin{document}

\title{Tunable circular dipole-like system in graphene: mixed electron-hole states}
\author{R. Van Pottelberge}\email{robbe.vanpottelberge@uantwerpen.be}
\affiliation{Departement Fysica, Universiteit Antwerpen, Groenenborgerlaan 171, B-2020 Antwerpen, Belgium}


\author{F. M. Peeters}\email{francois.peeters@uantwerpen.be}
\affiliation{Departement Fysica, Universiteit Antwerpen, Groenenborgerlaan 171, B-2020 Antwerpen, Belgium}
\begin{abstract}
Coupled electron-hole states are realised in a system consisting of a combination of an electrostatic potential barrier and ring shaped potential well, which resembles a circular dipole. A perpendicular magnetic field induces confined states inside the Landau gaps which are mainly located at the barrier or ring. Hybridization between the barrier and ring states are seen as anticrossings in the energy spectrum. As a consequence the energy levels show an oscillating dependence on the electrostatic potential strength in combination with an oscillating migration of the wave functions between the barrier and ring. At the anti-crossing points the quantum state consits of a mixture of electron and hole. The present system mimics closely the behaviour of a relativistic dipole on gapped graphene.  
\end{abstract}

\maketitle

\section{Introduction}
The experimental observation of graphene [\onlinecite{Novoselov}, \onlinecite{Geim}] as a stable 2D system together with its interesting electronic properties [\onlinecite{Neto}] has attracted a lot of attention. However, the gapless nature of the spectrum together with the linear spectrum has drastic consequences for the charge carriers in graphene: charge carriers cannot be confined by an electrostatic potential as shown by the Klein tunneling effect [\onlinecite{Katsnelson}]. Controlling the charge carriers in graphene is however essential for future electronic applications of graphene. 

One possibililty to control the charge carriers in graphene is by confining them in quantum dots. This can be done for example by cutting out a finite size flake of graphene which naturally confines the charge carriers [\onlinecite{Schnez}-\onlinecite{Chaves}]. However, it was shown in both theoretical and experimental studies that the nature of the edges of the finite size flakes drastically alters the energy spectrum [\onlinecite{Chaves}]. Since the edges are difficult to control experimentally this poses major challenges regarding future applications. 

However, in Refs. [\onlinecite{Giavaras}-\onlinecite{Giavaras2}] another possibility to control the charge carriers in graphene has been demonstrated. Here it was shown that by combining an electric and magnetic field a highly tunable quantum dot can be created. The magnetic field quantizes the energy spectrum and thus creates Landau gaps between the Landau levels. Using a nonhomogeneous electrical potential one is able to induce localized states that are inside those gaps. The high degree of tunablility of this type of dot system, together with the absence of edges makes it very promising for the use in future electronic applications of graphene (for example in quantum information and quantum computing [\onlinecite{Falko}]). Furthermore recent experiments [\onlinecite{Moriyama}-\onlinecite{He}] have demonstrated its high degree of tunability. 

\begin{figure}[h!!!!]
\includegraphics[trim={3cm 8cm 9cm 2cm},scale=0.60]{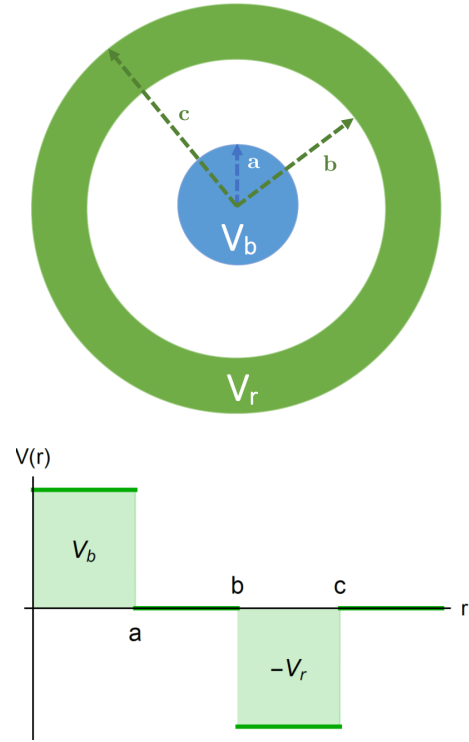}
\caption{Schematic representation of the dipole-like ring system, i.e. a circular potential barrier in the center of height $V_b$ and a ring potential well of depth $-V_r$. For the shape of the potential well and barriers we opt for constant piecewise step potentials which allow for analytical solutions. A magnetic field is applied surpressing the wave functions at larger distances.}
\end{figure}

Coupling between quantum dot states is important to understand because of their potential use for promising applications in quantum information and optoelectronics [\onlinecite{Doty}, \onlinecite{Zhang}]. Studying the coupling between two lateral quantum dots [\onlinecite{Apalkov}] is impossible analytically due to the breaking of angular symmetry. In this paper we consider an electrostatic potential barrier and ring (see Fig. 1) that are combined in a dipole-like configuration. By applying a perpendicular magnetic field it is possible to realize confinement in either the potential barrier and/or ring. We show that by tuning the magnetic and electric fields electron and hole states located at either the potential ring or barrier can be coupled. This coupling is highly tuneable by the external fields which paves the way towards the experimental study of magneto-electrostatic confined coupled graphene quantum dots. Furthermore we show that the general behaviour of the spectrum and probability densities closely mimics the behaviour of a relativistic dipole, hence our proposed system can provide a platform to study relativistic dipole physics.

By using step potentials and the circular symmetry of the system we are able to obtain formal analytical expressions for the energy spectrum and wave functions. Note that studying the interaction between two lateral quantum dots usually requires breaking angular symmetry which prevents exact analytical solutions.

The paper is structured as follows. In Sec. II we present the analytical model with expressions for the wave functions and the non linear equations that determine the energy spectrum. In order to understand the coupling between the circular quantum barrier and the ring shaped well we first consider both potential structures separately in Sec. III. The combination of a potential barrier and ring are studied in Sec IV. In Sec. V we present the main conclusions of this work. 

\section{Analytical Model}
In this section we present our system and derive the equations governing this system, the energy spectrum and wave functions.
\subsection{Model}
We consider a dipole-like structure created by a potential barrier surrounded by a potential well ring. We apply a magnetic field in order to create controllable confined states. A schematic representation of our system is given in Fig. 1. The potential barrier has a height of $V_b$ and radius $a$. It is surrounded by a ring shaped potential well with inner radius $b$, depth $V_r$ and width $c-b$. 

Depending on the strength of the magnetic field and electrostatic potentials interaction between wave functions located inside the ring and barrier will occur. In the next subsections we will present analytical details of the solutions. First we will consider the potential barrier and ring separately and subsequently we will couple the two systems. 

\subsection{Differential equations and solutions}
We will work within the continuum model. The Dirac-Weyl Hamiltonian for low-energy charge carriers in graphene, in the presence of a perpendicular magnetic field, is given by:
\begin{equation}
H=v_F(\bm{p}+e\bm{A})\cdot\bm{\sigma}+V(r)\cdot I.
\end{equation}
Here $V(r)$ is an arbitrary radially symmetric electrostatic potential, $\bm{\sigma}$ are the Pauli matrices and $\bm{A}=B/2(-y,x,0)$ is the vector potential in the symmetric gauge. In the present work we will consider $V(r)$ as being a step-wise potential.
The energy spectrum is determined by solving the Schr\"odinger equation $H\Psi=E\Psi$. 
Due to the circular symmetry we have $[H, J_z]=0$ where $J_z=L_z+\hbar\sigma_z /2$ is the total angular momentum. This implies that using polar coordinates the two-component wave function has the form $\Psi=(\phi_a,\phi_b)=e^{im\theta}(\psi_a(r),e^{i\theta}\psi_b(r))$, where $m=0,\pm 1,\pm 2,...$ is the total angular quantum number. Using the latter ansatz the coupled radial equations are given by:
\begin{subequations}
\begin{equation}
\frac{V(r)}{\hbar v_F}\psi_a+\left(\frac{\partial}{\partial r}+\frac{m+1}{r}\right)\psi_b+\frac{r}{2l_B^2}\psi_b=\frac{E}{\hbar v_F}\psi_a,
\label{eq:1}
\end{equation}
\begin{equation}
\label{eq:2}
\left(-\frac{\partial}{\partial r}+\frac{m}{r}\right)\psi_a+\frac{V(r)}{\hbar v_F}\psi_b+\frac{r}{2l_B^2}\psi_a=\frac{E}{\hbar v_F}\psi_b.
\end{equation}
\end{subequations}
Here $l_B=\sqrt{\hbar/eB}$ is the magnetic length and $E$ the energy. 
Solving Eq. (\ref{eq:1}) for $\psi_a$ gives the following expression in the case of a constant potential $V(r)=V$:
\begin{equation}
\label{eq:3}
\psi_a=\frac{\hbar v_F}{(E-V)}\left(\frac{\partial}{\partial r}+\frac{m+1}{r}\right)\psi_b+\frac{\hbar v_F}{(E-V)}\frac{r}{2l_B^2}\psi_b.
\end{equation}
Substituting the latter expression in Eq. (\ref{eq:2}) results in the uncoupled second order differential equation:
\begin{equation}
\begin{split}
\frac{\partial^2\psi_b}{\partial r^2}+\frac{1}{r}\frac{\partial \psi_b}{\partial r}-\\ \left[\frac{(m+1)^2}{r^2}+\frac{m}{l_B^2}+\frac{r^2}{4l_B^4}-\frac{(E-V)^2}{\hbar^2 v_F^2}\right]\psi_b=0.
\end{split}
\end{equation}
This equation has two independent exact solutions given by the following closed expressions, for the wave function:
\begin{subequations}
\begin{equation}
\label{eq:5}
\begin{split}
F_b(V,r)= & 2^{-\frac{m}{2}}e^{-r^2/4l_B^2}r^{-m-1} \\ &\times\text{L}\left[\frac{l_B^2}{2\hbar^2v_F^2}(E-V)^2,-1-m,\frac{r^2}{2l_B^2}\right],
\end{split}
\end{equation}
\begin{equation}
\label{eq:6}
\begin{split}
G_b(V,r)= &2^{-\frac{m}{2}}e^{-r^2/2l_B^2}r^{-m-1} \\& \times \text{U}\left[-\frac{l_B^2}{2\hbar^2v_F^2}(E-V)^2,-m,\frac{r^2}{2l_B^2}\right],
\end{split}
\end{equation}
\end{subequations}
where $L$ is the generalized Laguerre polynomial and $U$ is the Tricomi confluent hypergeometric function. The solution $F_b(V,r)$ is regular at the origin $r=0$ while it is irregular at infinity. The other solution $G_b(V,r)$ is irregular at the origin and regular at infinity.

The solutions for $\psi_a(r)$ can be obtained by plugging the solutions (\ref{eq:5}) and (\ref{eq:6}) into Eq. (\ref{eq:3}) giving the following wave functions:
\begin{subequations}
\begin{equation}
\label{eq:7}
\begin{split}
F_a(V,r)= & \frac{2^{-\frac{m}{2}}\hbar v_F}{l_B^2(E-V)}e^{-r^2/4l_B^2}r^{-m}\\ & \times \text{L}\left[-1+\frac{l_B^2}{\hbar^2 v_F^2}(E-V)^2,-m,\frac{r^2}{2l_B^2}\right],
\end{split}
\end{equation} 
and
\begin{equation}
\label{eq:8}
\begin{split}
G_a(V,r)= &\frac{2^{-\frac{m}{2}}(E-V)}{2\hbar v_F}e^{-r^2/4l_B^2}r^{-m}\\& \times\text{U}\left[1-\frac{l_B^2}{\hbar^2v_F^2}(E-V)^2,1-m,\frac{r^2}{2l_B^2
}\right].
\end{split}
\end{equation}
\end{subequations}
\subsection{Wave functions}
In this subsection we derive the wave functions for the potential barrier, potential ring and potential dipole system. 
\subsubsection{Potential barrier}
We consider first a circular potential barrier of radius $a$ and height $V_b$. Thus only two regions are relevant: in region I ($r<a$) the solution is given by the wave functions:
\begin{subequations}
\begin{equation}
\psi_a^{I}=\mathcal{A}_1F_a(V_b,r),
\label{eq:9}
\end{equation} 
\begin{equation}
\psi_b^{I}=\mathcal{A}_1F_b(V_b,r),
\label{eq:10}
\end{equation}
\end{subequations}
where $\mathcal{A}_1$ is an integration constant. 
In region II ($r>a$) we have to take the other solution for the wave functions
\begin{subequations}
\begin{equation}
\psi_a^{II}=\mathcal{B}_1G_a(0,r),
\label{eq:11}
\end{equation}
\begin{equation}
\psi_b^{II}=\mathcal{B}_1G_b(0,r),
\label{eq:12}
\end{equation}
\end{subequations}
where $\mathcal{B}_1$ is an integration constant. 

The above wave functions have to be matched at the boundary $r=a$, which results into the following wave functions in region I:
\begin{subequations}
\begin{equation}
\psi_a^{I}=\mathcal{B}_1\frac{G_a(0,a)}{F_a(V_b,a)}F_a(V_b,r),
\end{equation}
\begin{equation}
\psi_b^{I}=\mathcal{B}_1\frac{G_a(0,a)}{F_a(V_b,r)}F_b(V_b,r).
\end{equation}
\end{subequations}
In region II we have the following wave functions:
\begin{subequations}
\begin{equation}
\psi_a^{II}=\mathcal{B}_1G_a(0,r),
\end{equation}
\begin{equation}
\psi_b^{II}=\mathcal{B}_1G_b(0,r).
\end{equation}
\end{subequations}
\subsubsection{Potential ring}
Now we consider the situation where we have only a potential ring of depth $-V_r$ with inner(outer) radius b(c). In this case we have three regions. Region I is defined as $r<b$, where the wave function is given by the expressions:
\begin{subequations}
\begin{equation}
\psi_a^{I}=\mathcal{A}_2F_a(0,r),
\label{eq:13}
\end{equation}
\begin{equation}
\psi_b^{I}=\mathcal{A}_2F_b(0,r).
\label{eq:14}
\end{equation}
\end{subequations}
Region II is defined as $b<r<c$ where an electrostatic potential of strength $-V_r$ is present. Now we have to keep the two solutions of Eq. (1) and we have the following wave functions
\begin{subequations}
\begin{equation}
\psi_a^{II}=\mathcal{B}_2F_a(-V_r,r)+\mathcal{C}_2G_a(-V_r,r),
\label{eq:15}
\end{equation}
\begin{equation}
\psi_b^{II}=\mathcal{B}_2F_b(-V_r,r)+\mathcal{C}_2G_b(-V_r,r).
\label{eq:16}
\end{equation}
\end{subequations}
In region III, which is defined as $r>c$, we have the following wave functions:
\begin{subequations}
\begin{equation}
\psi_a^{III}=\mathcal{D}_2G_a(0,r),
\label{eq:17}
\end{equation}
\begin{equation}
\psi_b^{III}=\mathcal{D}_2G_b(0,r).
\label{eq:18}
\end{equation}
\end{subequations}
The wave functions have to be matched at $r=b$ and $r=c$. This leads to the wave functions given in appendix A. 
\subsubsection{Circular potential dipole}
In this situation we consider both a potential barrier and a ring, i.e. the combination of the previous two potential structures. In this case we have 4 regions. Region I is defined as $r<a$ with the wave functions
\begin{subequations}
\begin{equation}
\psi_a^{I}=\mathcal{A}_3F_a(V_b,r),
\label{eq:19}
\end{equation}
\begin{equation}
\psi_b^{II}=\mathcal{A}_3F_b(V_b,r).
\label{eq:20}
\end{equation}
\end{subequations}
In region II, where $a<r<b$, we have the following wave functions
\begin{subequations}
\begin{equation}
\psi_a^{II}=\mathcal{B}_3F_a(0,r)+\mathcal{C}G_a(0,r),
\label{eq:21}
\end{equation}
\begin{equation}
\psi_b^{II}=\mathcal{B}_3F_b(0,r)+\mathcal{C}G_b(0,r).
\label{eq:22}
\end{equation}
\end{subequations}
In region III, which is defined as $b<r<c$, we have an electrostatic potential $-V_r$, the corresponding wave functions are given by the expressions
\begin{subequations}
\begin{equation}
\psi_a^{III}=\mathcal{D}_3F_a(-V_r,r)+\mathcal{E}_3G_a(-V_r,r),
\label{eq:23}
\end{equation}
\begin{equation}
\psi_b^{III}=\mathcal{D}_3F_b(-V_r,r)+\mathcal{E}_3G_b(-V_r,r).
\label{eq:24}
\end{equation}
\end{subequations}
Last we have region IV which is defined as $r>c$, in this region we have the following wave functions
\begin{subequations}
\begin{equation}
\psi_a^{IV}=\mathcal{F}_3G_a(0,r),
\label{eq:25}
\end{equation}
\begin{equation}
\psi_b^{IV}=\mathcal{F}_3G_b(0,r).
\label{eq:26}
\end{equation}
\end{subequations}
Matching the wave functions at the boundaries $r=a$, $r=b$ and $r=c$ leads to the solutions which are given in appendix A. 

\subsection{Energy equations}
Using the wave functions derived in the previous subsection, we derive the equation for the energy spectrum. 
\subsubsection{Potential barrier}
The energy equation can be obtained by matching the wave functions (\ref{eq:9}) and (\ref{eq:10}) with the wave functions (\ref{eq:11}) and (\ref{eq:12}) at the boundary between the two regions. This gives the following algebraic equation:
\begin{equation}
\frac{F_a(V_b,a)}{F_b(V_b,a)}=\frac{G_a(0,a)}{G_b(0,a)},
\label{eq:27}
\end{equation}
whose solutions determine the energy spectrum. 

\subsubsection{Potential ring}
Matching the wave functions (\ref{eq:13}) and (\ref{eq:14}) with the wave functions (\ref{eq:15}) and (\ref{eq:16}) between region I and II gives the following relation
\begin{equation}
\frac{\psi_a^{I}}{\psi_b^{I}}=\frac{\psi_a^{II}}{\psi_b^{II}}\rightarrow \frac{F_a(0,b)}{F_b(0,b)}=\frac{\mathcal{B}_2F_a(-V_r,b)+\mathcal{C}_2G_a(-V_r,b)}{\mathcal{B}_2F_b(-V_r,b)+\mathcal{C}_2G_b(-V_r,b)},
\end{equation}
from which we obtain
\begin{equation}
\frac{\mathcal{B}_2}{\mathcal{
C}_2}=\frac{F_b(0,b)G_a(-V_r,b)-F_a(0,b)G_b(-V_r,b)}{F_a(0,b)F_b(-V_r,b)-F_b(0,b)F_a(-V_r,b)}.
\end{equation}
Matching the wave functions (\ref{eq:15}) and (\ref{eq:16}) with the wave functions (\ref{eq:17}) and (\ref{eq:18}) between region II and III gives the following relation
\begin{equation}
\frac{\psi_a^{III}}{\psi_b^{III}}=\frac{\psi_a^{II}}{\psi_b^{II}}\rightarrow \frac{G_a(0,c)}{G_b(0,c)}=\frac{\mathcal{B}_2F_a(-V_r,c)+\mathcal{C}_2G_a(-V_r,c)}{\mathcal{B}_2F_b(-V_r,c)+\mathcal{C}_2G_b(-V_r,c)},
\end{equation}
which results into the following equation:
\begin{equation}
\frac{G_a(0,c)}{G_b(0,c)}=\frac{\frac{\mathcal{B}_2}{\mathcal{C}_2}F_a(-V_r,c)+G_a(-V_r,c)}{\frac{\mathcal{B}_2}{\mathcal{C}_2}F_b(-V_r,c)+G_b(-V_r,c)},
\label{eq:31}
\end{equation}
from which we obtain the energy spectrum.

\subsubsection{Circular potential dipole}
The procedure is completely analogous as for the quantum barrier and quantum ring. 
Matching the wave functions (\ref{eq:19}) and (\ref{eq:20}) with the wave functions (\ref{eq:21}) and (\ref{eq:22}) between region I and II we can find the following ratio of integration constants
\begin{equation}
\frac{\mathcal{B}_3}{\mathcal{C}_3}=\frac{G_a(0,a)F_b(V_b,a)-G_b(0,a)F_a(V_b,a)}{F_b(0,a)F_a(V_b,a)-F_a(0,a)F_b(V_b,a)}.
\end{equation}
Matching the wave functions (\ref{eq:23}) and (\ref{eq:24}) with the wave functions (\ref{eq:25}) and (\ref{eq:26}) between region I and IV we find the following ratio
\begin{equation}
\frac{\mathcal{D}_3}{\mathcal{E}_3}=\frac{G_b(-V_r,c)G_a(0,c)-G_a(-V_r,c)G_b(0,c)}{F_a(-V_r,c)G_b(0,c)-F_b(-V_r,c)G_a(0,c)}.
\end{equation}
Finally, matching the wave functions (\ref{eq:21}) and (\ref{eq:22}) with the wave functions (\ref{eq:23}) and (\ref{eq:24}) between region II and III we obtained the following equation
\begin{equation}
\frac{F_a(0,b)+\frac{\mathcal{C}_3}{\mathcal{B}_3}G_a(0,b)}{F_b(0,b)+\frac{\mathcal{C}_3}{\mathcal{D}_3}G_b(0,b)}=\frac{F_a(-V_r,b)+\frac{\mathcal{E}_3}{\mathcal{D}_3}G_a(-V_r,b)}{F_b(-V_r,b)+\frac{\mathcal{E}_3}{\mathcal{B}_3}G_b(-V_r,b)},
\label{eq:34}
\end{equation}
whose solutions give the energy spectrum. 

\begin{figure*}[t]
\includegraphics[scale=1.14,trim={1cm 0.5cm 1cm 0.5cm}]{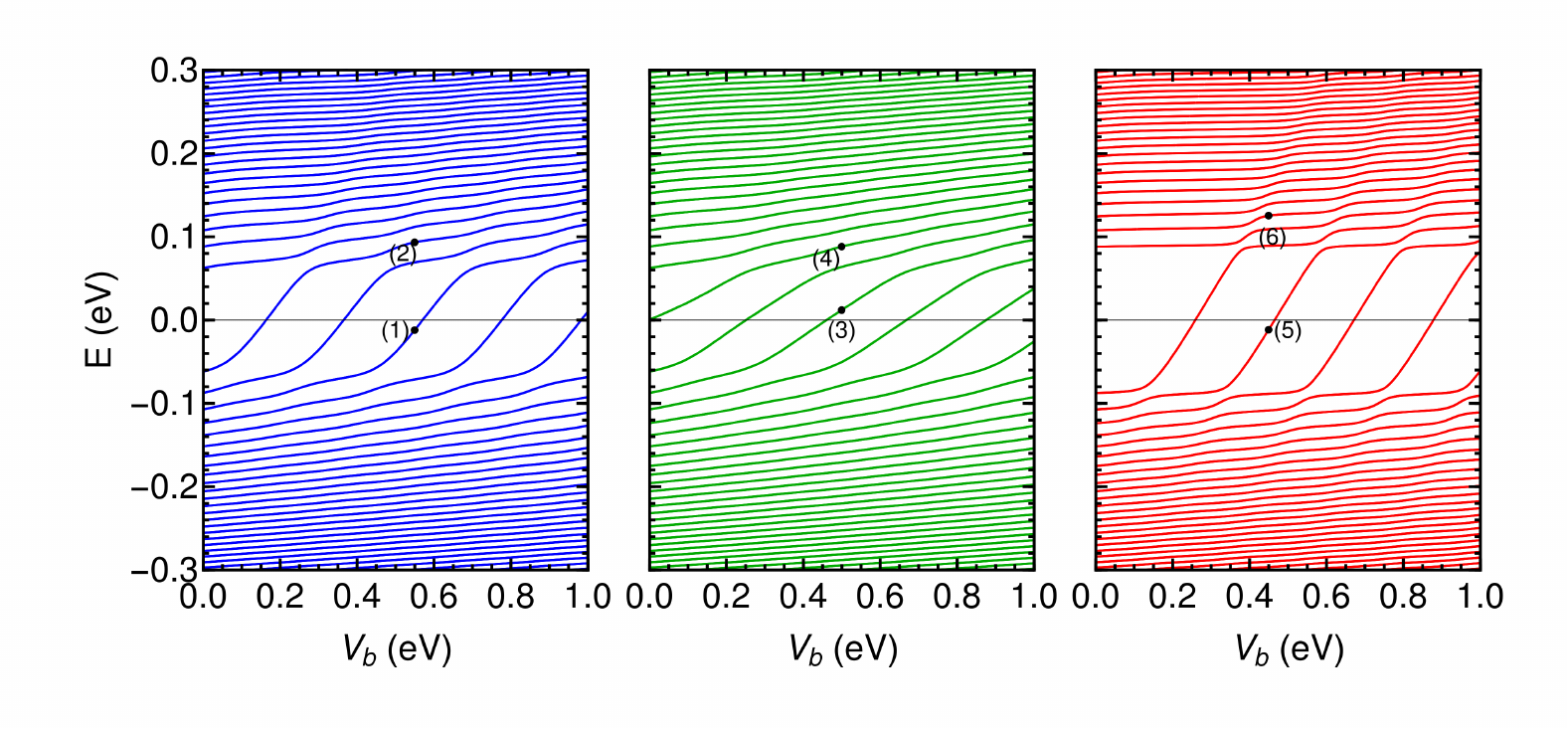}
\caption{Energy spectrum of a circular potential barrier of radius $a=10$ nm as function of the potential barrier strength for three values of the angular momentum quantum number $m=0$ (blue curves), $m=-1$ (green curves) and $m=1$ (red curves) and in the presence of a perpendicular magnetic field of $B\approx 3$ T corresponding to a magnetic length $l_B=15$ nm.}
\end{figure*}

\begin{figure}[h!!!!!!!!!!!]
\includegraphics[scale=0.29]{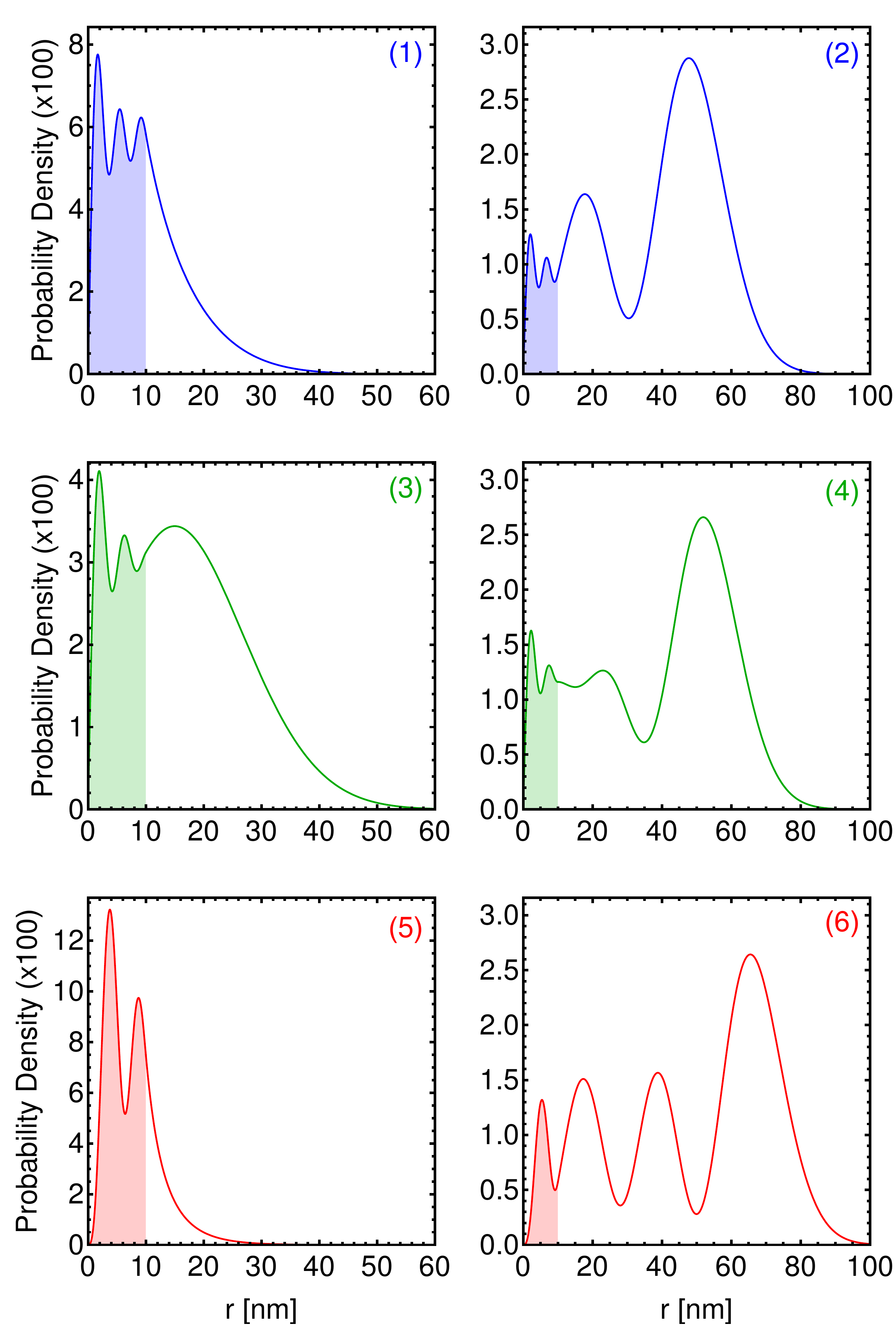}
\caption{Probability density shown for the points marked in Fig. 2. The position of the potential barrier is marked by the colored surface under the density profile. The left side panels correspond to the density of states inside the Landau gap while the right panels correspond to states outside the Landau gap.}
\end{figure}
\section{Numerical results for the decoupled systems}
In this section we will discuss the results obtained from solving the energy equations (\ref{eq:27}) and (\ref{eq:31}). 

\subsection{Potential barrier}
We consider first the results for the simplest system consisting of just a circular quantum barrier and solve numerically the energy equation (\ref{eq:27}) for different values of the potential strength and angular quantum number. This system was investigated previously in Ref. [\onlinecite{Giavaras}] but to understand the spectrum of a dot-ring system we repeat here the essential results. 

In Fig. 1 we show the energy spectrum as function of the potential barrier strength $V_b$ for three angular quantum number values $m=0,\pm 1$ and a magnetic length $l_B=15$ nm (corresponding to $B\approx 3$ T). We took the strength of the dot potential positive, effectively creating a potential barrier. However, the spectrum is symmetric in the sense that $V_b\rightarrow -V_b$ is equivalent to $E\rightarrow-E$. This means that changing the sign of the barrier corresponds to interchanging electron and hole states. 
 
For the quantum numbers shown ($m=0,1,-1$) in Fig. 2 we see Landau gaps. These gaps are determined by the Landau levels $E_n=\pm\hbar v_F/l_B\sqrt{2\mid N\mid}$ and consequently decrease with decreasing magnetic field. When the potential barrier strength increases hole states rise into the Landau gap region and form quantum dot states. In the Landau gaps hole states are allowed to rise further towards the more slowly rising electron Landau levels outside the gap region with increasing potential strength. This continues untill at some point the first state inside the gap region reaches the first electron state and anticrosses with the corresponding electron level. This behavior reminds of the supercritical instability effect in gapped graphene where bound states in the gap are allowed to enter the corresponding continuum [\onlinecite{peeters}-\onlinecite{pereira}], with the important difference that in that case the effect is created by a Coulomb impurity and not a potential barrier. The fact that a potenential of arbitrary shape could be used to create supercritical states was shown explicitely in Ref. [\onlinecite{Aoki}].

In order to show the difference in behaviour for the states inside the Landau gap as compared to those outside it we show in Fig. 3 the probability densities for the points marked in Fig. 2. The location of the barrier is shown as a colored area under the probability density plots. From these figures it is clear that the states inside the Landau gap (left figures in Fig. 3) are more localized inside the potential barrier and form true quantum dot states. States outside the gap region are weakly localised in the potential barrier and these states exhibit more a Landau level like behaviour. Interestingly for the states located in the gap region we find that the $m=-1$ state (green curve) is much less localised in the barrier as compared to the $m=0$ (blue curve) and $m=1$ (red curve) state. This explains the smoother dependence of the energy, compared to the $m=1$ states, as function of the barrier strength. 

\begin{figure*}[t]
\includegraphics[scale=1.14, trim={1cm 0.5cm 1cm 0.5cm}]{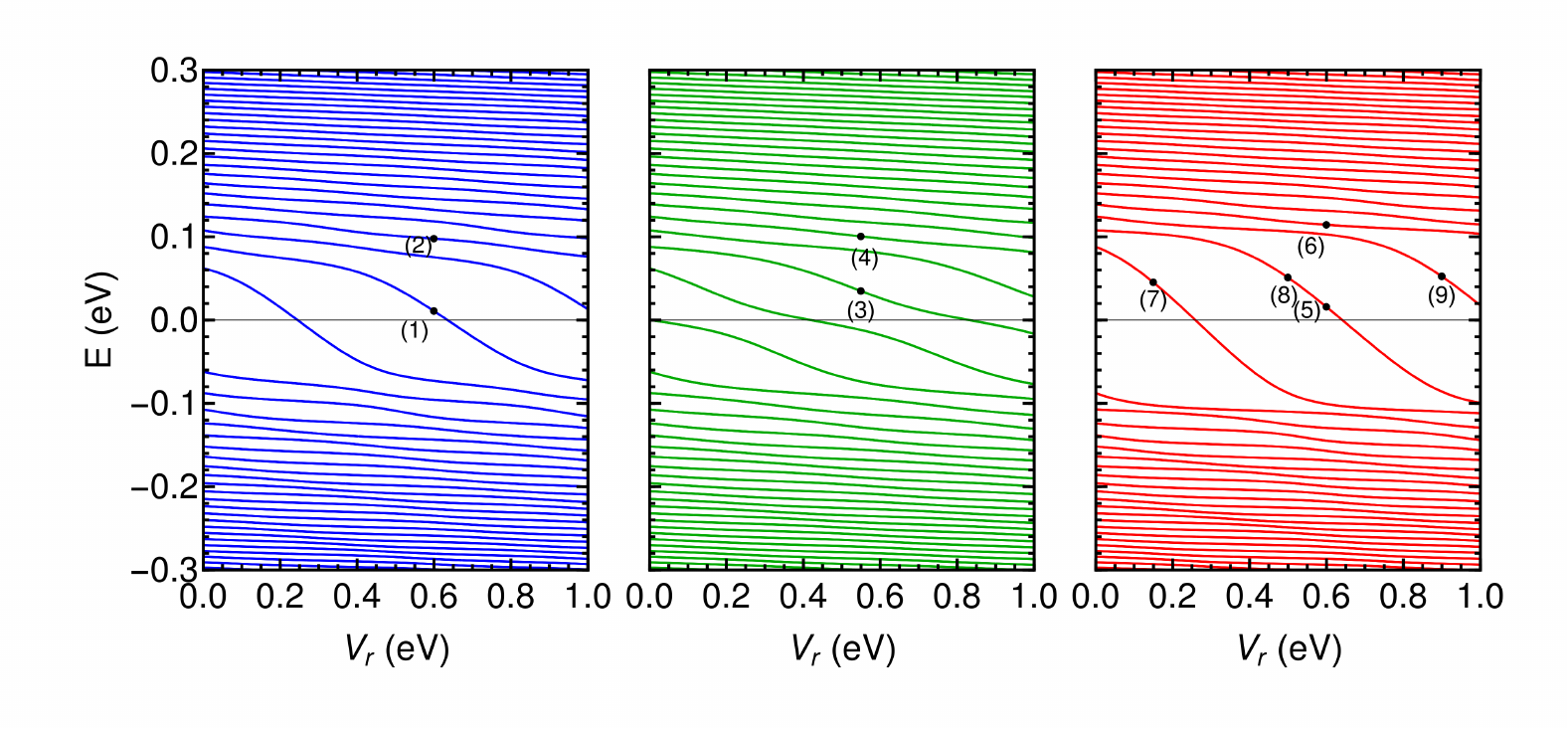}
\caption{Energy spectrum of a ring potential as function of the depth of the potential well for three values of the angular momentum quantum number $m=0$ (blue curves), $m=-1$ (green curves) and $m=1$ (red curves) with inner(outer) radius $b=30$ nm($c=35$ nm). A perpendicular magnetic field is applied with strength $B\approx3$ T which results in $l_B=15$ nm.}
\end{figure*}

\begin{figure}
\includegraphics[scale=0.30]{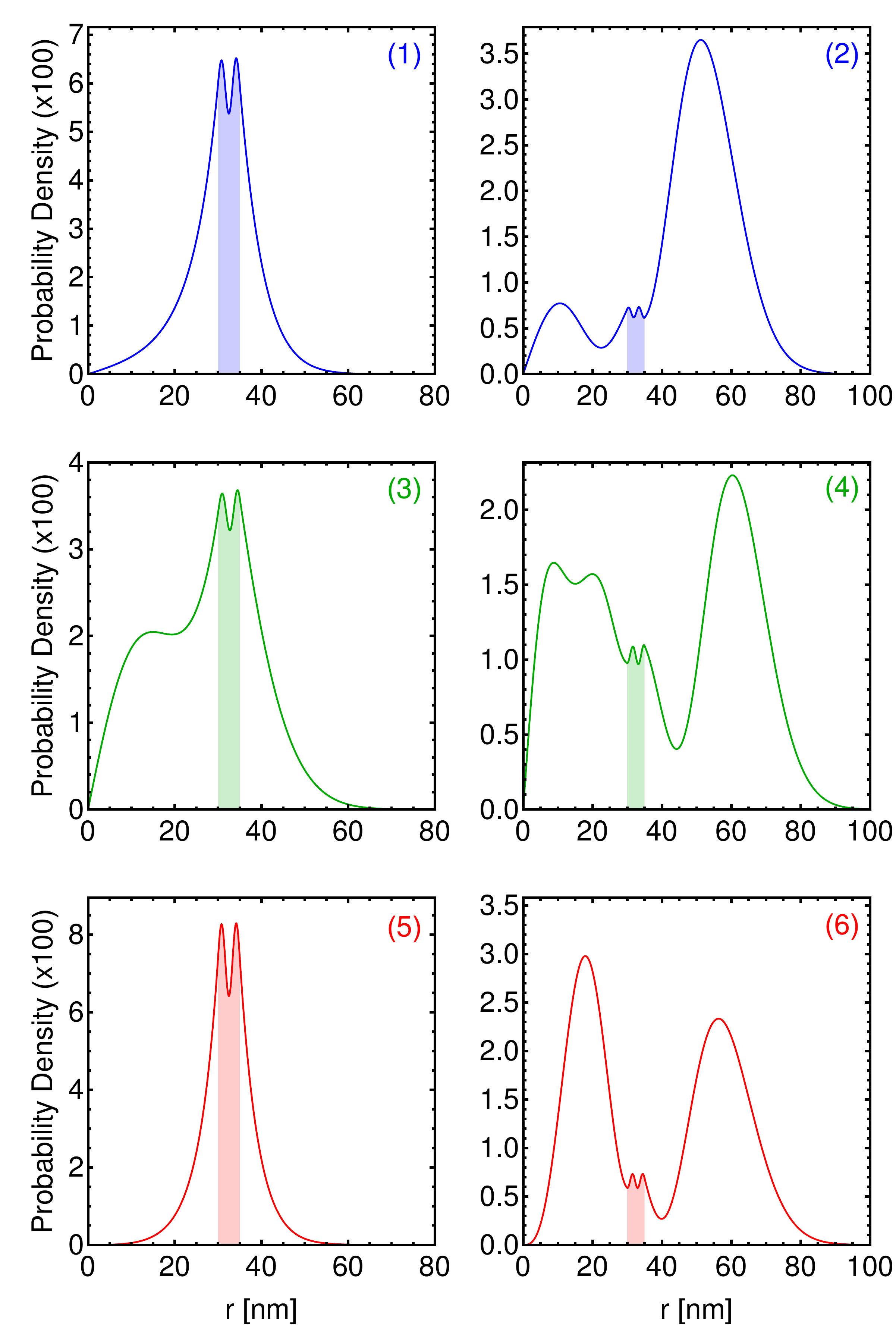}
\caption{Probability density shown for the points (1)-(6) marked in Fig. 4. The area with the potential ring is marked by the colored surface under the density profile. The left side panels correspond to the probability density of states inside the Landau gap while right panels correspond to the states outside the Landau gap.}
\end{figure}

\subsection{Potential ring}
Now we consider an electrostatic potential ring and investigate the corresponding energy spectrum. In the presence of a perpendicular magnetic field it is possible to create quantum dot states located in the circular quantum ring within the Landau gaps. We emphasize the differences between quantum dot states located at the barrier compared to those located at a ring shaped potential. While graphene quantum rings cut out from graphene sheets have been extensively studied both within tight binding and continuum models [\onlinecite{Trauzettel}-\onlinecite{Chaves2}], the system discussed in this paper, the combination of a ring shaped potential and magnetic field, has not been studied in the literature to date. 

In Fig. 4 we show the spectrum of an electrostatic quantum well ring for the same three angular quantum numbers as shown in Fig. 2 for the potential barrier. For the ring the size parameters are $b=30$ nm and $c=35$ nm.

\begin{figure}
\includegraphics[scale=0.42]{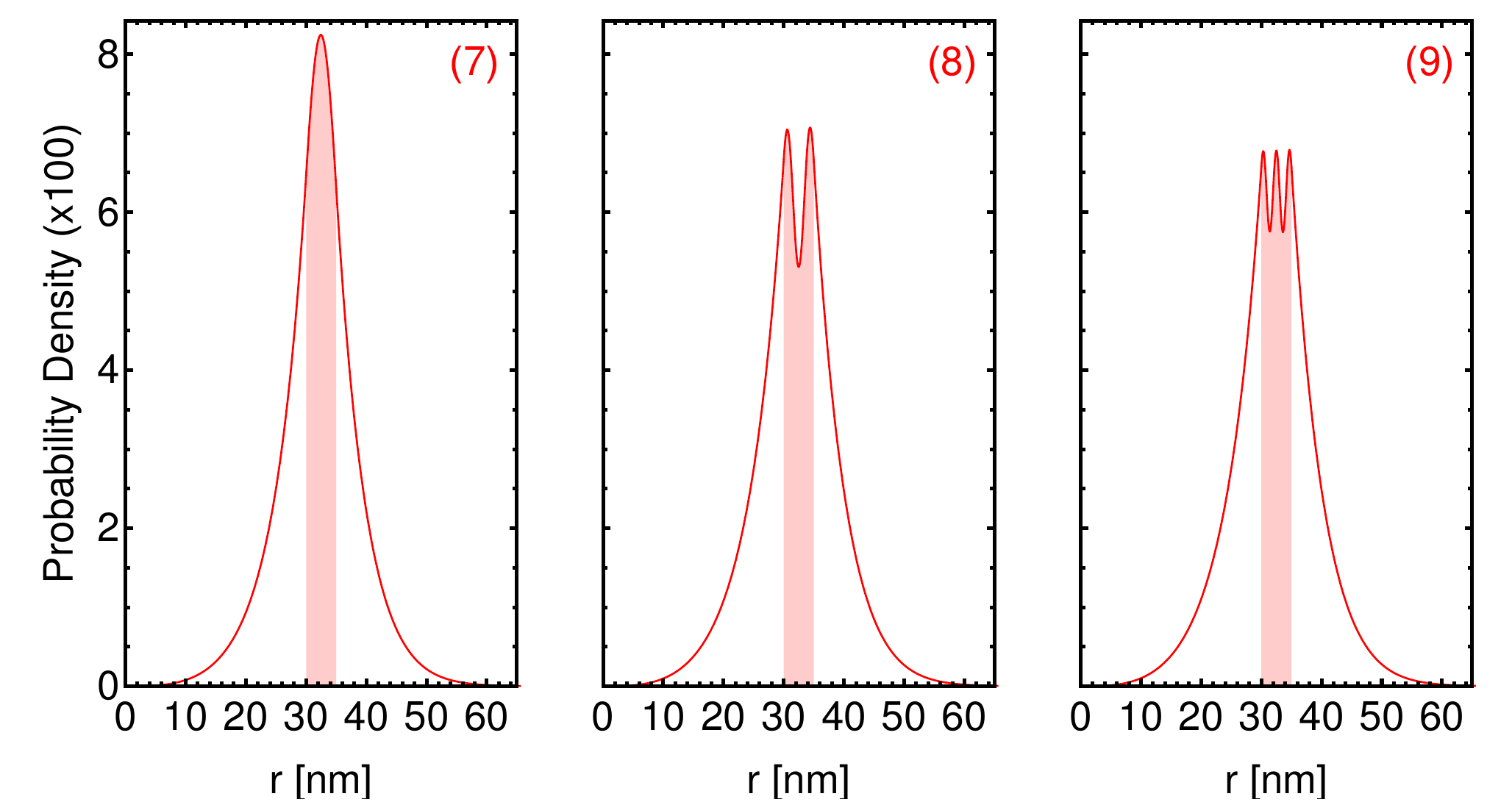}
\caption{Probability densities for the points (7)-(9) marked in Fig. 4. The area with the potential ring is indicated by the colored surface under the probability densities.}
\end{figure}

\begin{figure*}[t]
\includegraphics[scale=1.14, trim={1cm 0cm 0cm 0cm}]{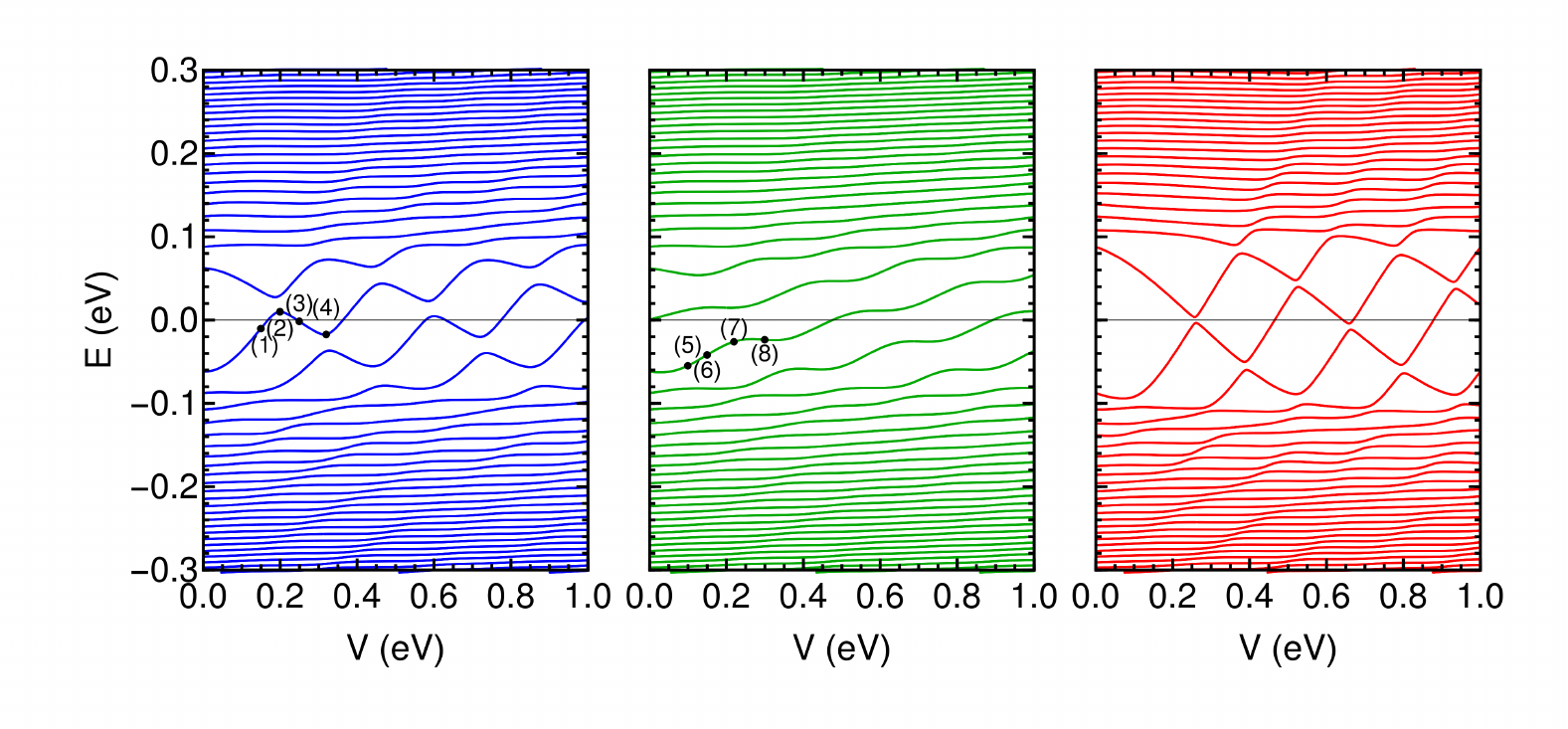}
\caption{Energy spectrum as function of the potential ring and barrier strength $V=V_r=V_b$ for three values of the angular momentum quantum number $m=0$ (blue curves), $m=-1$ (green curves) and $m=1$ (red curves). The size of the dipole system we took $a=10$ nm, $b=30$ nm and $c=35$ nm and the magnetic field value is $l_B\approx 3$ T corresponding to $l_B=15$ nm.}

\end{figure*}
In contrast with the potential barrier no hole states enter the gap region but electron states enter instead. This is merely a consequence of the fact that we took a potential well ring instead of a barrier. Remember that the spectrum has electron-hole symmetry in the sense that the spectrum is invariant under the transformations $E\rightarrow -E$ and $V_r\rightarrow -V_r$, effectively changing electron states into hole ones. With increasing potential strength electron states are allowed to sink into the Landau gaps where they exhibit almost a linear dependence on the potential strength. This behaviour continues until the electron level touches the first hole level at negative energy and an anticrossing occurs, similar to the above quantum barrier results.


In Figure 5 the probability density is shown for the points (1)-(6) marked in Fig. 4. The left panels show the densities for points located inside the Landau gap region while the right panels show densities for states located outside the Landau gap. In the left panels it can be nicely seen how the probability densities are located around the quantum ring. As in the case of a potential barrier it is seen that the $m=-1$ are less localised around the quantum ring compared to the $m=0$ and $m=-1$ states. This explains the smoother behaviour on the potential ring strength of the $m=-1$ states compared to the other angular momenta states. For the energy states located outside the gap region (right panels of Fig. 5) the behaviour is totally different. These states do not exhibit a large peak in the probability density at the potential ring and are not localised in the ring. These states behave almost as unperturbed Landau levels which is reflected in the behaviour of the probability densities which show a Landau Level like behavior. Note that regardless of the fact that these states are clearly less localized at the quantum ring they still feel the potential ring. This can be seen from the fact that in the right panels of Fig. 5 small sub-peaks are observed in the probability density located inside the quantum ring. This is a manifestation of the Klein tunnelling providing a coupling between the states outside the gap and the potential ring. 

In Fig. 6 the probability densities are shown for the three successive states in the gap region shown in Fig. 4 for the angular quantum number $m=1$ and $l_B=15$ nm, indicated by the points (7)-(9). As explained in the previous paragraph the densities are clearly spatially localized in the quantum ring. In general for all three states the densities exhibit similar behaviour. However, the number of peaks inside the electrostatic ring increases with the number of states entering the gap region.  The first state has one peak, the second two and the third three. This behaviour reminds of the increasing number of nodes with increasing principal quantum number in the case of a relativistic hydrogen atom. Interestingly the behavior outside the quantum ring is almost exactly the same for the three successive states and these states are only distinguished by the behaviour inside the quantum ring. 

\section{Circular potential dipole}
Here, we investigate the interaction between the single electron energy spectrum of a combined quantum barrier and ring shaped well. By combining a barrier and ring with opposite sign for electrostatic potential strength it is possible to couple electron and hole states which shows up as anticrossings. This coupling can be effectively tuned by the strength of the applied electrostatic and magnetic field. 

In Fig. 7 we show the spectrum as function of the electrostatic potential strength for three values of the angular momentum. We use the same dimensions for the barrier and ring as used in the previous sections, i.e. $a= 10$ nm, $b=30$ nm and $c=35$ nm and the same value for the magnetic length ($l_B=15$ nm). From Fig. 7 it is clear how electron states descend from the upper continuum while hole states rise from the lower continuum and enter the Landau gaps where they approach each other and anti-cross. These anticrossings become stronger when the magnetic field is further reduced, thus the interaction between the barrier and rings states (i.e. the strength of the anticrossings) can be effectively tuned by the magnetic field. Note that anticrossings are only observed for states with the same angular momentum quantum number, which is a consequence of the Wigner-Von Neumann theorem [\onlinecite{Wigner}]. With increasing electrostatic field strength more states anticross and the states inside the Landau gap start to show an oscillating dependence on the electrostatic potential strength.

In Fig. 8 we show the probability densities of the points (1)-(4) shown in Fig. 7 which are the electronic states with $m=0$ around an anticrossing region. Before the anticrossing the electronic state is mainly located at the potential barrier and is mostly hole, which explains the fact that the energy is increasing with the electrostatic potential strength. At the point of anticrossing however the state is spread out over both the barrier and the ring and represents a coupled electron-hole state. This shows that at the point of anticrossing a hybridization between a state from the quantum barrier and quantum ring occurs. After the anticrossing the probability moves entirely to the quantum ring and the state becomes electron like. At the next point of anticrossing the state is again evenly distributed over the quantum barrier and ring. 

\begin{figure}
\includegraphics[scale=0.35]{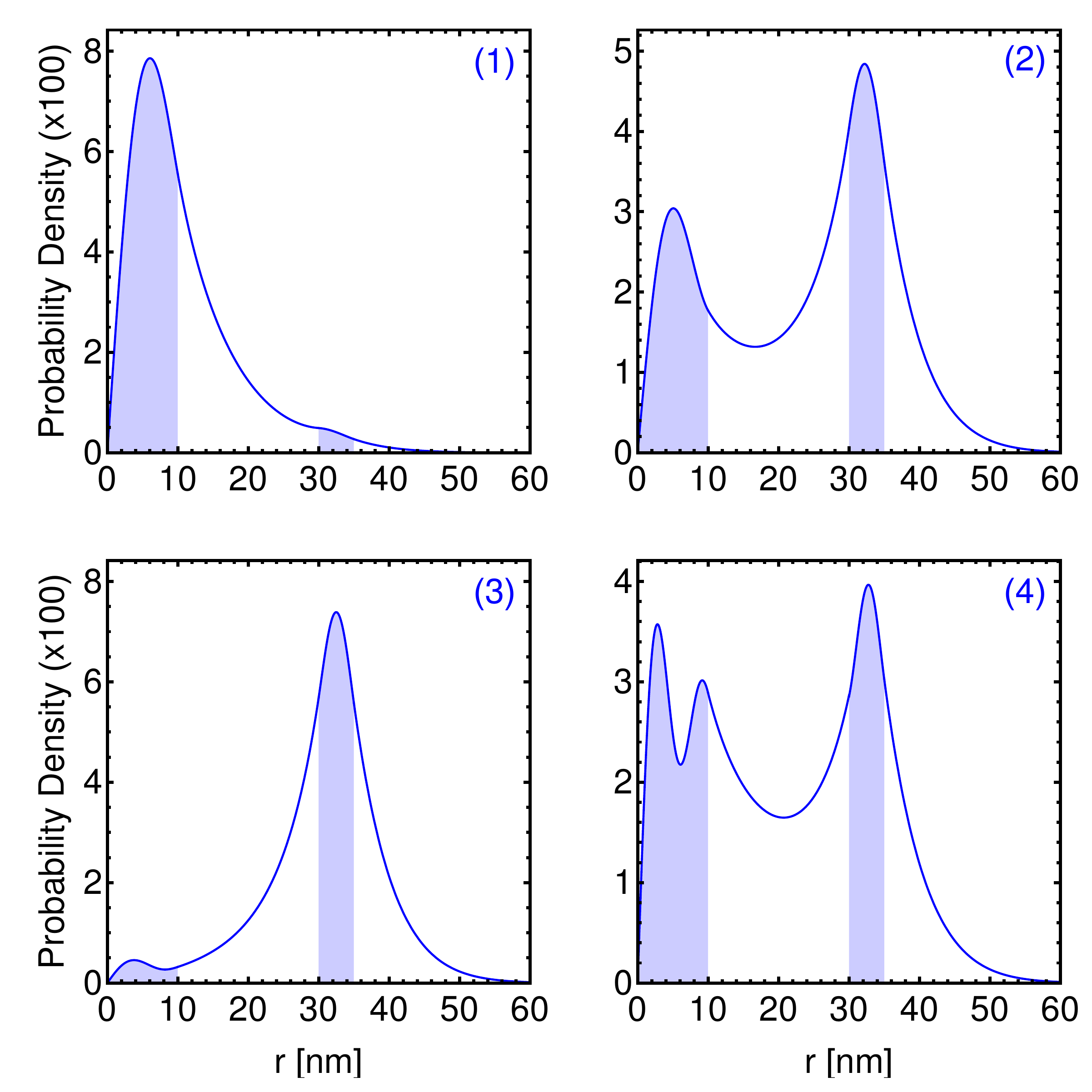}
\caption{Probability densities for the points shown in the $m=0$ energy spectrum of Fig. 7.}
\end{figure}
\begin{figure}
\includegraphics[scale=0.35]{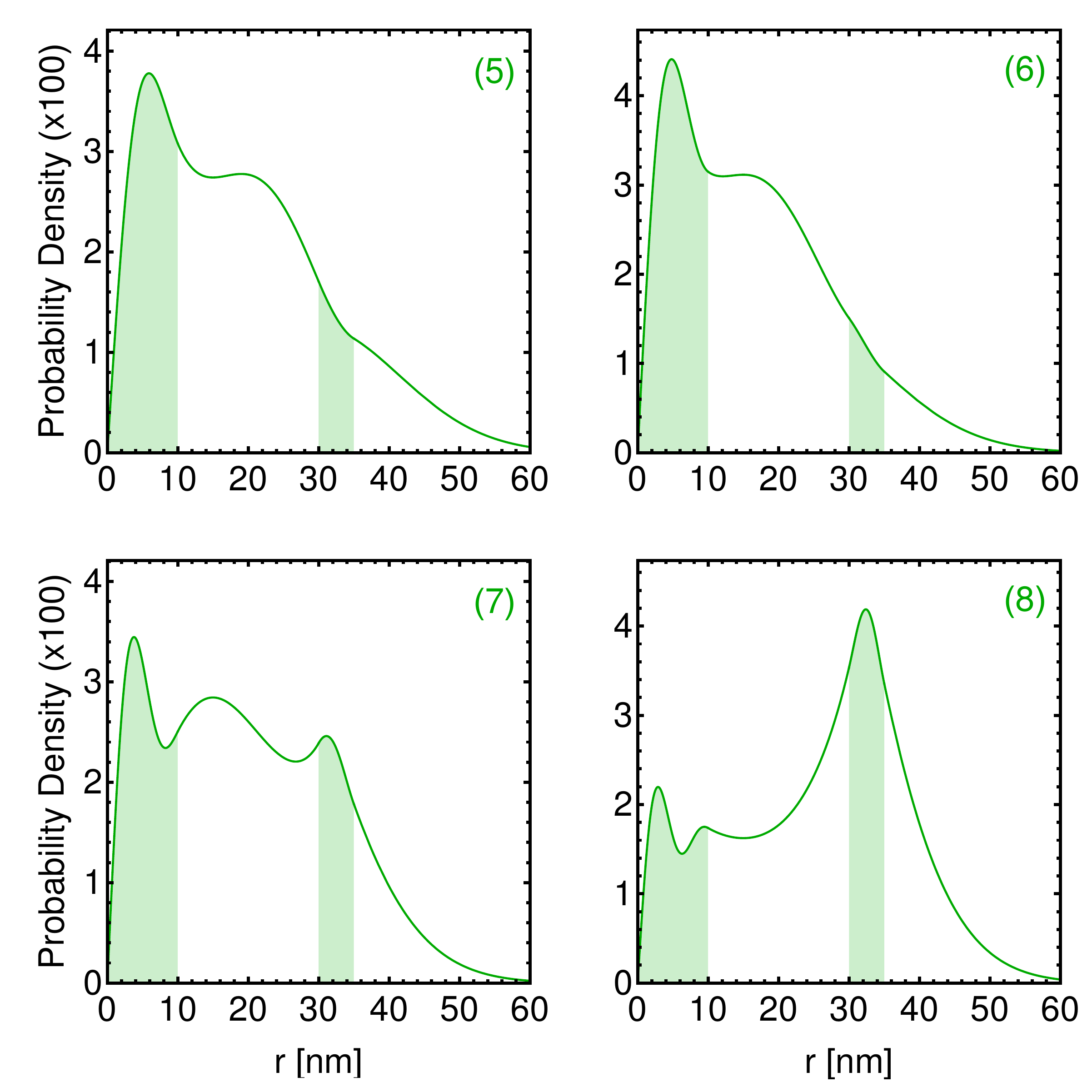}
\caption{Probability densities for the points shown in  $m=-1$ energy spectrum of Fig. 7.}
\end{figure}

From Fig. 7 we notice that the anticrossings are less clear for the $m=-1$ states as compared to the other angular momenta values. This is a consequence of the fact that the states are less localized inside the barrier and/or ring for $m=-1$ (as discussed in the previous sections). This argument is supported by Fig. 9 where we show the probability densities for the points (5)-(8) shown in Fig. 7. Qualitatively the behavior is very similar to that of the densities shown in Fig. 8, the localisation shifts from the barrier to the ring and is equally distributed at the point of anticrossing. However, quantitatively before and after the point of anticrossing the probability density is more spread out over the barrier and ring as compared to the densities shown in Fig. 8. 

Interestingly the behaviour shown in this paper, i.e. the oscillations of the energy spectrum as function of the electrostatic potential strength and relocalisation of the wave function when passing the anticrossing, are also observed in the spectrum of a relativistic dipole on gapped graphene [\onlinecite{dipole1}-\onlinecite{dipole4}]. Thus the coupled quantum barrier and ring system in a magnetic field imitates relativistic dipole physics in gapped graphene. Both systems share some similarities, the magnetic field creates a gap in the spectrum while the electrostatic barrier and ring which are equal in strength but opposite in sign replicate a positively and negatively charged Coulomb impurity. However, experimentally the system presented in this paper has several advantages as compared to the dipole system in gapped graphene. First, the present system has many more tuning possibilities as compared to the dipole system, hence increasing the chance of detecting the rich physics of a dipole. The governing length scales can be effectively tuned by the magnetic field while the strength of the electrostatic potentials are tunable by nanostructured gates. Second, analytical results are obtained in our paper which was impossible for a dipole on gapped graphene. 

\section{Conclusions}
In this paper we presented a system consisting of a potential barrier and potential ring, e.g. a model system for a circular dipole. In the first part of the paper we derived the analytical expressions for the wave functions and energy equations for the potential barrier and potential ring separately and when they are both present. In the second part of the paper we solved numerically the energy equations in order to determine the energy spectra and wave functions.

We showed that in the case of a potential barrier and potential ring states enter the Landau gaps, created by the magnetic field, and become spatially localized at the potential barrier and ring. We studied the spectrum and wave functions for different potential ring and barrier strengths, different values of the magnetic field and different values of the angular momentum quantum number $m$. 

By combining a potential barrier and potential ring equal in strength but opposite in sign, i.e. a dipole like structure, we showed that electron states originating from the potential ring and hole states originating from the potential barrier are allowed to hybridize and form coupled states, which are seen in the spectrum as anticrossings in the Landau gaps. The coupling between the quantum states and hence the strength of the anticrossings can be effectively tuned by e.g. the magnetic field and the strength of the electrostatic potentials. Finally we showed that due to the similarities with a relativistic dipole placed on gapped graphene the states for a dipole system mimic closely the behaviour of the electronic states of a dipole on gapped graphene.

Our model has the big advantage that analytical results for the energy equations and wave functions can be obtained, the energy equations provided in this paper can be straightforwardly solved using standard root solving methods. However, one should always consider a trade-off between simplicity of the model and experimental relevance. In this paper we consider step potentials which in experiments can be approximately realized by an STM tip, or by local doping, or by nanostructured gates. In real experiments the potentials produced in this way will deviate from these step potentials. However, a previous publication [\onlinecite{Giavaras}] has shown that the use of step potentials makes sense and that they provide a good approximation for the potentials present in real systems.  

\acknowledgements

We thank Matthias Van der Donck for fruitful discussions. This work was supported by the Research Foundation of Flanders (FWO-V1) through an aspirant research grant for RVP. 
\newline

\appendix
\section{Wave functions}
In this appendix we provide the exact form of the wave functions obtained from our anlytical model. 
\subsection{Potential ring}
In region I we have the wave functions:
\begin{equation}
\psi_a^{I}=\mathcal{A}_2F_a(0,r),
\end{equation}
\begin{equation}
\psi_b^{I}=\mathcal{A}_2F_b(0,r).
\end{equation}
In region II we have the wave functions:
\begin{equation}
\psi_a^{II}=\mathcal{A}_2.\mathbb{B}\left(\mathbb{A}F_a(-V_r,r)+G_a(-V_r,r)\right),
\end{equation}
\begin{equation}
\psi_b^{II}=\mathcal{A}_2.\mathbb{B}\left(\mathbb{A}F_b(-V_r,r)+G_b(-V_r,r)\right).
\end{equation}
Here $\mathbb{A}$ is defined as
\begin{equation}
\mathbb{A}=\frac{G_a(-V_r,c)G_b(0,c)-G_b(-V_r,c)G_a(0,c)}{F_b(-V_r,c)G_a(0,c)-F_a(-V_r,c)G_b(0,c)},
\end{equation}
and $\mathbb{B}$ is defined as
\begin{equation}
\mathbb{B}=\frac{F_a(0,b)}{\mathbb{A}F_a(-V_r,b)+G_a(-V_r,b)}.
\end{equation}
In region III we have the following wave functions:
\begin{equation}
\psi_a^{III}=\mathcal{A}_2\mathbb{C}G_a(0,r),
\end{equation}
\begin{equation}
\psi_b^{III}=\mathcal{A}_2\mathbb{C}G_b(0,r).
\end{equation}
Here $\mathbb{C}$ is given by the following expression
\begin{equation}
\mathbb{C}=\frac{\mathbb{B}\left(\mathbb{A}F_a(-V_r,c)+G_a(-V_r,c)\right)}{G_a(0,c)}.
\end{equation}
\subsection{Circular potential dipole}
In region I we have the wave functions:
\begin{equation}
\psi_a^{I}=\mathcal{A}_3F_a(V_b,r),
\end{equation}
\begin{equation}
\psi_b^{I}=\mathcal{A}_3F_b(V_b,r).
\end{equation}
In region II we have the wave functions:
\begin{equation}
\psi_a^{II}=\mathcal{A}_3\mathbb{F}\left(\mathbb{D}F_a(0,r)+G_a(0,r)\right),
\end{equation}
\begin{equation}
\psi_b^{II}=\mathcal{A}_3\mathbb{F}\left(\mathbb{D}F_b(0,r)+G_b(0,r)\right).
\end{equation}
Here $\mathcal{D}$ and $\mathcal{F}$ are respectively given by the expression
\begin{equation}
\mathbb{D}=\frac{G_a(0,a)F_b(V_b,a)-G_b(0,a)F_a(V_b,a)}{F_b(0,a)F_a(V_b,a)-F_a(0,a)F_b(V_b,a)}
\end{equation}
and
\begin{equation}
\mathbb{F}=\frac{F_a(V_b,a)}{\mathbb{D}F_a(0,a)+G_a(0,a)}.
\end{equation}
In region III we have the following wave functions
\begin{equation}
\psi_a^{III}=\mathcal{A}_3\mathbb{G}\left(\mathbb{E}F_a(-V_r,r)+G_a(-V_r,r)\right),
\end{equation}
\begin{equation}
\psi_b^{III}=\mathcal{A}_3\mathbb{G}\left(\mathbb{E}F_b(-V_r,r)+G_b(-V_r,r)\right).
\end{equation}
Here $\mathcal{E}$ and $\mathcal{G}$ are respectively given by the expressions
\begin{equation}
\mathbb{E}=\frac{G_a(-V_r,c)G_b(0,c)-G_b(-V_r,c)G_a(0,c)}{F_b(-V_r,c)G_a(0,c)-F_a(-V_r,c)G_b(0,c)}
\end{equation}
and
\begin{equation}
\mathbb{G}=\frac{\mathbb{F}\left(\mathbb{D}F_a(0,b)+G_a(0,b)\right)}{\mathbb{E}F_a(-V_r,b)+G_a(-V_r,b)}.
\end{equation}
In region IV we have the wave functions:
\begin{equation}
\psi_a^{IV}=\mathcal{A}_3\mathbb{H}G_a(0,r),
\end{equation}
\begin{equation}
\psi_b^{IV}=\mathcal{A}_3\mathbb{H}G_b(0,r).
\end{equation}
Here $\mathbb{H}$ is given by the expression
\begin{equation}
\mathbb{H}=\frac{\mathbb{G}\left(\mathbb{E}F_a(-V_r,c)+G_a(-V_r,c)\right)}{G_a(0,c)}.
\end{equation}

\end{document}